%
\let\includefigures=\iffalse
%
\let\useblackboard=\iftrue
%
%
\newfam\black
\input harvmac.tex
\input diagrams.tex
\input rotate
\input epsf
\noblackbox
\includefigures
\message{If you do not have epsf.tex (to include figures),}
\message{change the option at the top of the tex file.}
\def\figin{\epsfcheck\figin}\def\figins{\epsfcheck\figins}
\def\epsfcheck{\ifx\epsfbox\UnDeFiNeD
\message{(NO epsf.tex, FIGURES WILL BE IGNORED)}
\gdef\figin##1{\vskip2in}\gdef\figins##1{\hskip.5in}
\else\message{(FIGURES WILL BE INCLUDED)}%
\gdef\figin##1{##1}\gdef\figins##1{##1}\fi}
\def\DefWarn#1{}
\def\figinsert{\goodbreak\midinsert}
\def\ifig#1#2#3{\DefWarn#1\xdef#1{fig.~\the\figno}
\writedef{#1\leftbracket fig.\noexpand~\the\figno}%
\figinsert\figin{\centerline{#3}}\medskip\centerline{\vbox{\baselineskip12pt
\advance\hsize by -1truein\noindent\footnotefont{\bf Fig.~\the\figno:} #2}}
\bigskip\endinsert\global\advance\figno by1}
\else
\def\ifig#1#2#3{\xdef#1{fig.~\the\figno}
\writedef{#1\leftbracket fig.\noexpand~\the\figno}%
\global\advance\figno by1}
\fi
\useblackboard
\message{If you do not have msbm (blackboard bold) fonts,}
\message{change the option at the top of the tex file.}
\font\blackboard=msbm10
\font\blackboards=msbm7
\font\blackboardss=msbm5
\textfont\black=\blackboard
\scriptfont\black=\blackboards
\scriptscriptfont\black=\blackboardss
\def\Bbb#1{{\fam\black\relax#1}}
\else
\def\Bbb{\bf}
\fi
%
\def\yboxit#1#2{\vbox{\hrule height #1 \hbox{\vrule width #1
\vbox{#2}\vrule width #1 }\hrule height #1 }}
\def\fillbox#1{\hbox to #1{\vbox to #1{\vfil}\hfil}}
\def\ybox{{\lower 1.3pt \yboxit{0.4pt}{\fillbox{8pt}}\hskip-0.2pt}}

\def\mapr{\mathop{\longrightarrow}\limits}
\def\rightarrowbox#1#2{
  \setbox1=\hbox{\kern#1{${ #2}$}\kern#1}
  \,\vbox{\offinterlineskip\hbox to\wd1{\hfil\copy1\hfil}
    \kern 3pt\hbox to\wd1{\rightarrowfill}}}

\def\comments#1{}

\def\QC{\Bbb{C}}

\def\QZ{\Bbb{Z}}
\def\p{\partial}

\def\tr{{\rm tr\ }}

\def\lcm{{\rm lcm}}

\def\bra#1{{\langle}#1|}
\def\ket#1{|#1\rangle}

\def\vev#1{\langle{#1}\rangle}

\def\CN{{\cal N}}
\def\CO{{\cal O}}

\def\mapr{\mathop{\longrightarrow}\limits}

\def\P{\BP}
\def\WP{\BW\BP}
\def\I{I}

\def\II{\relax{I\kern-.10em I}}
\def\IIa{{\II}a}

\def\IZ{\relax\ifmmode\mathchoice
{\hbox{\cmss Z\kern-.4em Z}}{\hbox{\cmss Z\kern-.4em Z}}
{\lower.9pt\hbox{\cmsss Z\kern-.4em Z}}
{\lower1.2pt\hbox{\cmsss Z\kern-.4em Z}}\else{\cmss Z\kern-.4em
Z}\fi}
\def\IB{\relax{\rm I\kern-.18em B}}
\def\IC{{\relax\hbox{$\inbar\kern-.3em{\rm C}$}}}
\def\ID{\relax{\rm I\kern-.18em D}}
\def\IE{\relax{\rm I\kern-.18em E}}
\def\IF{\relax{\rm I\kern-.18em F}}
\def\IG{\relax\hbox{$\inbar\kern-.3em{\rm G}$}}
\def\IGa{\relax\hbox{${\rm I}\kern-.18em\Gamma$}}
\def\IH{\relax{\rm I\kern-.18em H}}
\def\II{\relax{\rm I\kern-.18em I}}
\def\IK{\relax{\rm I\kern-.18em K}}
\def\IN{\relax{\rm I\kern-.18em N}}
\def\IP{\relax{\rm I\kern-.18em P}}
\def\IW{{\bf W}}

\def\inbar{\,\vrule height1.5ex width.4pt depth0pt}

\def\dgam{{d|\Gamma|}}
\def\p{\partial}

\font\cmss=cmss10 \font\cmsss=cmss10 at 7pt
\def\IR{\relax{\rm I\kern-.18em R}}

\def\Hom{{\rm Hom}}

\def\Pic{{\rm Pic}}

\def\ch{{\hbox{ch}}}
\def\td{{\hbox{Td}}}

\def\BR{\IR}
\def\BZ{\QZ} 
\def\BP{\IP}
\def\BW{\IW}
\def\BR{\IR}
\def\BC{\QC}

\def\lp10{l_P^{10}}
\def\lp11{l_P^{11}}
\def\R11{R_{11}}

\newbox\tmpbox\setbox\tmpbox\hbox{\abstractfont IASSNS-HEP-00/50, 
RUNHETC-2000-27}
\Title{\vbox{\baselineskip12pt\hbox to\wd\tmpbox{\hss
hep-th/0006224}\hbox{IASSNS-HEP-00/50, RUNHETC-2000-27}}}
{\vbox{
\centerline{D-branes on Stringy Calabi--Yau Manifolds}}}
\smallskip
\centerline{Duiliu-Emanuel Diaconescu$^1$ and 
Michael R. Douglas$^{2,}$\footnote{$^{3}$}{
Louis Michel Professor}}
\bigskip
\bigskip
\medskip
\centerline{$^1$School of Natural Sciences, Institute for Advanced Study, Princeton, NJ 08540 USA}
\medskip
\centerline{$^2$Department of Physics and Astronomy, Rutgers University, Piscataway, NJ 08855-0849 USA}
\medskip
\centerline{$^3$I.H.E.S., Le Bois-Marie, Bures-sur-Yvette, 91440 France}
\bigskip
\centerline{\tt diacones@ias.edu, mrd@physics.rutgers.edu}
\bigskip

\bigskip
\noindent
We argue that D-branes corresponding to rational B boundary states in
a Gepner model can be understood as fractional branes in the Landau--Ginzburg
orbifold phase of the linear sigma model description.
Combining this idea with the generalized McKay correspondence allows
us to identify these states with coherent sheaves, and 
to calculate their K-theory classes in the large
volume limit, without needing to invoke mirror symmetry.
We check this identification against the mirror symmetry results
for the example of the Calabi--Yau hypersurface in $\WP^{1,1,2,2,2}$.

\Date{June 2000}
\def\np{{\it Nucl. Phys.}}

\nref\partres{C. Beasley, B. R. Greene, C. I. Lazaroiu, and 
  M. R. Plesser, ``D3-branes on partial resolutions of abelian 
  quotient singularities of Calabi--Yau threefolds'', {\it Nucl. Phys.}
  {\bf B566} (2000) 599, hep-th/9907186.}
\nref\beilinson{A. A. Beilinson, ``Coherent sheaves on $\IP^n$ and
  problems of linear algebra'', {\it Funct. Anal. Appl.}
  {\bf 12} (1978) 214--216.}
\nref\bkr{T. Bridgeland, A. King, and M. Reid, ``Mukai implies McKay'', 
  math.AG/9908027.}
\nref\bdlr{I. Brunner, M. R. Douglas, A. Lawrence, and C. R\"omelsberger,
  ``D-branes on the quintic,'' hep-th/9906200.}
\nref\ilka{I. Brunner and V. Schomerus, to appear.}
\nref\twopar{P. Candelas, X. de la Ossa, A. Font, S. Katz, and 
  D. R. Morrison,
  ``Mirror symmetry for two parameter models -- I'', 
  {\it Nucl. Phys.} {\bf B416} (1994) 481, hep-th/9308083.}
\nref\candelas{P. Candelas, X. C. de la Ossa, P. S. Green, and 
  L. Parkes, ``A pair of
  {C}alabi--{Y}au manifolds as an exactly soluble superconformal theory'',
  \np\ {\bf B359} (1991) 21.}
\nref\cox{D. Cox, ``Recent developments in toric geometry'', {\it Proc. Symp.
  Pure Math.} {\bf 62:2} (1997) 389--436, alg-geom/9606016.}
\nref\inprogress{D.-E. Diaconescu and M. R. Douglas, work in progress.}
\nref\dg{D.-E. Diaconescu and J. Gomis, ``Fractional branes and boundary
  states in orbifold theories,'' hep-th/9906242.}
\nref\dr{D.-E. Diaconescu and C. R\"omelsberger, ``D-branes and 
  bundles on elliptic fibrations'', {\it Nucl. Phys.} {\bf B574} (2000) 
  245, hep-th/9910172.}%
\nref\dtopics{M. R. Douglas, ``Topics in D-geometry'',
  {\it Class. Quant. Grav.} {\bf 17} (2000) 1057, hep-th/9910170.}
\nref\category{M. R. Douglas, ``D-branes and categories'', to appear.}
\nref\pistable{M. R. Douglas, B. Fiol, and C. R\"omelsberger,
  ``Stability and BPS branes'', hep-th/0002037.}
\nref\noncompact{M. R. Douglas, B. Fiol, and C. R\"omelsberger, 
  ``The spectrum of BPS branes on a noncompact Calabi--Yau'', 
  hep-th/0003263.}
\nref\dgm{M. R. Douglas, B. R. Greene, and D. R. Morrison, ``Orbifold
  resolution by D-branes,'' {\it Nucl. Phys.} {\bf B506} (1997) 84,
  hep-th/9704151.}
\nref\dm{M. R. Douglas and G. Moore, 
  ``D-branes, quivers, and ALE instantons,''
  hep-th/9603167.}
\nref\fulton{W. Fulton, {\it Introduction to toric varieties}, 
  Annals of Mathematics Studies, vol.~131, The William H. Roever Lectures in 
  Geometry, Princeton University Press, 1993.}
\nref\govj{S. Govindarajan and T. Jayaraman, 
  ``On the Landau--Ginzburg description
  of boundary CFTs and special Lagrangian submanifolds'', 
  hep-th/0003242.}
\nref\govjs{S. Govindarajan, T. Jayaraman, and T. Sarkar, 
  ``World sheet approaches to D-branes 
  on supersymmetric cycles'', hep-th/9907131.}
\nref\vafa{K. Hori, A. Iqbal, and C. Vafa, 
  ``D-branes and mirror symmetry'', hep-th/0005247.}
\nref\hulltown{C. M. Hull and P. K. Townsend, ``Enhanced gauge symmetries
  in superstring theories'', {\it Nucl. Phys.} {\bf B451} (1995) 
  525.}
\nref\itonakajima{Y. Ito and H. Nakajima, ``McKay correspondence and
  Hilbert schemes in dimension three'', math.AG/9803120.}
\nref\johnsonmyers{C. V. Johnson and R. C. Meyers, ``Aspects of type 
  IIB theory on ALE spaces'', {\it Phys. Rev.} {\bf D55} (1997) 6382, 
  hep-th/9610140.}
\nref\lerche{P. Kaste, W. Lerche, C. A. L\"utken, and J. Walcher,
  ``D-branes on K3 fibrations,'' hep-th/9912147.}
\nref\kronone{P. B. Kronheimer, ``The construction of ALE spaces as 
  hyper-K\"ahler quotients'', {\it J. Differential Geom.} {\bf 29} (1989)
  665.}
\nref\krontwo{P. B. Kronheimer,
  ``A Torelli-type theorem for gravitational instantons'',
  {\it J. Differential Geom.} {\bf 29} (1989) 685.}
\nref\kronnak{P. B. Kronheimer and H. Nakajima, 
  ``Yang--Mills instantons on ALE gravitational instantons'', 
  {\it Math. Ann.} {\bf 288} (1990) 263.}
\nref\mckay{J. McKay, ``Graphs, singularities, and finite groups'', 
  {\it Proc. Symp. in Pure Math.} {\bf 37} (1980) 183.}
\nref\MM{R. Minasian and G. Moore, 
``K-theory and Ramond-Ramond charge'', {\it JHEP} {\bf 11} (1997) 002,
hep-th/9710230.}
\nref\quantcoh{
  D. R. Morrison and M. R. Plesser, 
  `` Summing the instantons: Quantum cohomology and mirror symmetry 
  in toric varieties'',  {\it Nucl. Phys.} {\bf B440} (1995) 279,
  hep-th/9412236.}
\nref\naknozak{M. Naka and M. Nozaki, 
  ``Boundary states in Gepner models,'' {\it JHEP}\/ 0005 (2000) 027,
  hep-th/0001037.}
\nref\rs{A. Recknagel and V. Schomerus, ``D-branes in Gepner models'',
  \np\ {\bf B531} (1998) 185, hep-th/9712186.}
\nref\reid{M. Reid, ``McKay correspondence'', alg-geom/9702016.}
\nref\reidbourbaki{M. Reid, ``La correspondance de McKay'', S\'eminaire
  Bourbaki (novembre 1999), no. 867, math.AG/9911165.}
\nref\infirri{A. V. Sardo Infirri, ``Partial resolutions of orbifold
  singularities via moduli spaces of HYM-type bundles,'' alg-geom/9610004.}
\nref\infirritwo{A. V. Sardo Infirri, ``Resolutions of orbifold 
  singularities and flows on the McKay quiver,'' alg-geom/9610005.}
\nref\schei{E. Scheidegger, 
  ``D-branes on some one- and two-parameter Calabi--Yau hypersurfaces,'' 
  {\it JHEP}\/ 0004 (2000) 003, hep-th/9912188.}
\nref\warner{N.P. Warner,
  ``Supersymmetry in Boundary Integrable Models,''
  {\it Nucl. Phys.} B450 (1995) 663-694;
hep-th/9506064.}
\nref\witten{E. Witten, ``Phases of $N{=}2$ theories in two dimensions,''
  {\it Nucl. Phys.} {\bf B403} (1993) 159, hep-th/9301042.}
\nref\DK{E. Witten,``D-Branes and K-Theory'', {\it JHEP} 
{\bf 12} (1998) 019,  hep-th/9810188.}
%
%
\newsec{Introduction}

D-branes in Calabi--Yau compactification of string theory have been
the focus of a number of recent works.
In this work we continue the study of D-branes at Gepner points initiated in
\refs{\rs,\bdlr}.  We will show how many results for the spectrum of
rational boundary states and the corresponding brane world-volume
theories can be derived starting from the linear sigma model.  The
basic idea is to realize the boundary states as fractional branes in
the Landau--Ginzburg orbifold phase; we will show how recent
mathematical work on the generalized McKay correspondence determines
the identification of these boundary states as bundles in the large
volume limit, and check this identification in an example against results
obtained using mirror symmetry.  As in \noncompact, this framework allows
identifying bound states of branes with bundles and provides explicit 
descriptions of their moduli spaces; we will pursue this in more detail
in subsequent work.

For an overview of this line of work, we refer to \dtopics.
The starting point is Gepner's identification
of certain $\CN=2$ CFT's as stringy Calabi--Yau manifolds (CYs).
In \rs, rational boundary states (those which can be
easily obtained as orbifold products of boundary states in the individual
$\CN=2$ minimal models) were constructed for Gepner models. 
This provides explicit CFT realizations of D-branes on these manifolds,
and allows computing RR charges (in a natural basis at the Gepner point),
as well as the number of marginal operators.  It is also possible to compute
superpotentials, as outlined in \bdlr\ and as has been done in examples 
\ilka.

The natural extension of Gepner's identification would be to identify
these BPS boundary states with specific D-branes in the large volume
limit of the same Calabi--Yau.  
The work \bdlr\ made first steps towards such an identification.
The ``decoupling conjecture'' made there gives strong reasons to think
that B branes at any point in K\"ahler moduli space
should be identifiable with specific holomorphic objects
(bundles, coherent sheaves or complexes) in the large volume limit.
Using a derivation of the K\"ahler moduli space
from mirror symmetry \candelas, an explicit translation
of the RR charges of the B boundary states in the $(3)^5$ Gepner model
into Chern classes was made, which determines the topological type of
the corresponding bundles in the large volume limit.  Similar results
for other Calabi--Yau manifolds have been obtained in
\refs{\dg,\lerche,\schei,\naknozak}.

In \noncompact\ the $\BC^3/\BZ_3$ orbifold was studied in detail, and a
remarkable relation was found between the quiver gauge theory of
\refs{\infirri,\infirritwo,\dgm} 
and Beilinson's construction  \beilinson\ of holomorphic vector
bundles on $\P^2$: the quiver theory and mirror symmetry
results reproduce this construction, providing a very detailed
correspondence between F-flat configurations of the gauge theory and
holomorphic bundles in the large volume limit.
It was also pointed out that the
results of \bdlr\ for the quintic had a very similar relationship to
Beilinson's construction of bundles on $\P^4$.

The present work will explain and generalize this relationship.
Besides the work above, it is inspired by a generalization of
Beilinson's construction developed in recent mathematical work on
the generalized McKay correspondence 
\refs{\reid,\itonakajima,\bkr}.

The basic idea is to realize the Calabi--Yau threefold of interest as a
submanifold of the resolution of a higher dimensional orbifold
$\BC^n/\Gamma$, define D-branes in the higher dimensional orbifold
using the construction of Douglas and Moore \dm, and then identify the
D-branes of interest as the restriction of these to the original CY.
As we explain, this procedure can also be directly motivated by the
physics of boundary states in the linear sigma model construction of
the CY \refs{\witten,\govjs,\govj,\vafa}.

{\it Acknowledgements.}  We would like to express our special thanks to
Dave Morrison for collaboration on the early stages of this project, 
and for many valuable discussions and suggestions.

\newsec{Gepner models and quivers}

\subsec{Gepner models and linear sigma models}

A Gepner model is a product of $r$ minimal
models at level $k_i$ whose central charges $3k_i/(k_i+2)$ add to $3n$.
As we will review shortly, this corresponds to a Fermat hypersurface in a weighted
projective space, which if $n+r$ is even is $\WP(w_i)$, 
where $w_i=K/(k_i+2)$ and
$K=\lcm \{k_i+2\}$.  If $n+r$ is odd, we adjoin $w_{r+1}=K/2$ to this
list, and henceforth take $n+r$ even.  One can show that for $r=n+2$, these
requirements imply that $K=\sum w_i$.

When $n=3$, such a Gepner model can also be realized as a $(2,2)$ linear sigma
model \witten.  It has a $U(1)$ gauge group, $r$ chiral superfields
$Z^i$ with charges $w_i$, and a chiral superfield $P$ with charge
$-K$.  There is a superpotential $W_G = P \sum_i (Z^i)^{k_i+2}$.  The
D-flatness conditions are
\eqn\lindflat{
\zeta = \sum_i w_i |Z^i|^2 - K |P|^2
} with an FI parameter $\zeta$, and the model has two phases depending
on this parameter.  The Gepner model is associated with the
``Landau--Ginzburg'' phase with $\zeta<0$ and $\vev{P}\ne 0$; the
action expanded around this configuration is a sum of $\CN=2$
Landau--Ginzburg models, while the $U(1)$ symmetry is broken to
$\BZ_K$.
On the other hand, $\zeta>0$ produces the ``geometric'' phase in which D-flat
configurations with $P=0$
parameterize the weighted projective space $\WP(w_i)$, while the condition 
$0=\p W_G/\p P$ defines the CY as a hypersurface in this space.

It will be useful to have a picture of the space of D-flat
configurations (in other words, the vacua of the corresponding theory
with no superpotential) in the two phases.  The general D-flat
configuration in the geometric phase allows $P\ne 0$ and the total
space is a line bundle over $\WP(w_i)$.  For $K=\sum w_i$ this is the
anticanonical bundle, and the total space is itself a Calabi--Yau,
generically singular because of the singularities of $\WP(w_i)$.  Let
us denote this Calabi--Yau as $X(w_i)$ or simply $X$.

Similarly, in the Landau--Ginzburg phase the general D-flat
configuration has $Z^i\ne 0$.  The condition \lindflat\ determines $P$
and since $\vev{P}\ne 0$ always, the $U(1)$ gauge symmetry is always
broken to $\BZ_K$.  Thus the D-flat moduli space in this phase is
$\BC^r/\BZ_K$ with the $\BZ_K$ action defined by the action of a
generator
$$g( Z^i ) = e^{2\pi i w_i/K} Z^i.$$ 
Note that this generates a
discrete subgroup of $SU(r)$, so this noncompact orbifold is also a
CY.

In a later section, we will review the toric description of these
configuration spaces and the relation between these two phases.  The
general idea is that the algebra of holomorphic functions on the
configuration space is independent of the D-flatness conditions, and
thus must be the same in the two phases.  Thus we can consider the
space $X(w_i)$ as a (partial) resolution of the noncompact orbifold
$\BC^r/\BZ_K$.  In both phases, the superpotential will confine the
theory to the CY$_3$ as a hypersurface in the exceptional divisor,
$Z^i=0$ in $\BC^r/\BZ_K$, and the resolution of this point
$\pi^{-1}(0)$ in $X$.

\subsec{B boundary states}

Our basic claim is that the rational B boundary states can be thought
of as the restriction of the ``fractional brane'' states of the
$\BC^r/\BZ_K$ orbifold to the CY$_3$.

A fractional brane state in a $\BC^r/\Gamma$ orbifold is a Dirichlet
boundary state in $\BC^r$, with an additional choice of an irreducible
representation of $\Gamma$.  A collection of fractional branes is
labeled by a representation $R$ of $\Gamma$ or equivalently the
multiplicities $n_a$ of the irreps $\gamma_a$.  The world-volume
theory of the collection is then derived from the world-volume theory
of $\dim R$ branes in $\BC^r$ by projecting on invariants under the
action of $\Gamma$ on the fields twisted by the $\gamma_a$'s; in
particular vectors $Z^i$ in $\BC^r$ (such as those parameterizing
transverse motion of the branes) are projected as
$$\gamma_R(g)^{-1} Z^i \gamma_R(g) = (\gamma_{def})^i_j(g) Z^j.$$

These definitions do not require $\hat c\le 10$ or even that the bulk
theory of interest be a conformal field theory.  Thus we can apply
them directly to the LG orbifold phase of the linear sigma model.  It
is known that Dirichlet boundary conditions are $\CN=2$ supersymmetric
in the ungauged LG model \refs{\warner,\govjs,\vafa}.  If we work far
below the scale of $U(1)$ gauge symmetry breaking (set by the vev of
$P$ and thus the FI term), the full (B type) linear sigma model boundary
conditions must reduce to conventional Dirichlet boundary conditions for
$Z^i$.  We need only take the unbroken discrete gauge symmetry into
account, which is what is done by the fractional brane prescription.

Now, since we start with a non conformal theory, we must expect the IR
spectrum of marginal operators to be rather different from the UV free
theory spectrum, raising the question of what world-volume theory we
should take for the branes.

We do know that the flow must preserve the massless Ramond states, as
these are protected by the usual index considerations.  We can thus
compute the massless Ramond spectrum in the UV and carry it to the IR.

We then make the crucial assumption that---although the combinations
of BPS branes we are considering together break all supersymmetry
(they preserve different $\CN=1$ subalgebras of the original 
$\CN=2$)---this supersymmetry breaking is a spontaneous supersymmetry 
breaking
in an effective $\CN=1$, $d=4$ world-volume theory.  In particular,
combinations of BPS branes which together would break supersymmetry
can lead to BPS bound states, which are simply described by
(quasi)-supersymmetric vacua of the combined theory.  This assumption
is not completely obvious, especially as we will be discussing fields
with string scale masses in the broken supersymmetry vacua, and as we
will see it is literally true only for a subset of the theories.  It
is further discussed and motivated in
\category.  In any case, we proceed to postulate an effective $\CN=1$
world-volume theory which is compatible with our information.

The massless open string Ramond sector for the CFT of $r$ free
superfields will simply be a spinor (of definite chirality) of
$SO(2r)$, or equivalently a sum of antisymmetric representations of
$SU(r)$.  In the familiar case of $\BC^3$ orbifolds, this leads to a
singlet and a vector of $SU(3)$, and supersymmetry incorporates these
into space-time vector and chiral multiplets respectively.  The
resulting world-volume theory is the familiar $\CN=4$ super Yang--Mills
and its dimensional reductions, to which the $\Gamma$ projection is
applied.

In the case of $\BC^5$ orbifolds, these considerations lead to a
vector of $SU(5)$, a three index antisymmetric tensor, and a singlet
(equivalently a five index antisymmetric tensor).  We then assume that
the flow to the IR leads to a $(2,2)$ supersymmetric theory with an
$\CN=1$, $d=4$ world-volume interpretation.  On general grounds, the
nonsinglets will have to enter into chiral multiplets in this theory.
The singlet might enter into either chiral or vector multiplets {\it a
priori}, but given that any boundary theory will contain the operator
$1$ which is the internal CFT part of the gauge boson vertex operator,
there must be a vector multiplet in the space-time theory, whose
fermion must be this singlet.

This motivates the claim that the world-volume theory of $N$ D-branes
on $\BZ^5/\Gamma$ (assuming $\CN=1$ supersymmetry) is a $U(N)$ gauge
theory with $15$ chiral multiplets in the adjoint of $U(N)$,
transforming as ${\bf 5}+{\overline{\bf 10}}$ 
of a global $SU(5)$.  Let us denote these
multiplets as $X^i$ and $Y^{[ij]}$ respectively.

Such a theory admits gauge invariant superpotentials, and the leading
possible term is cubic:
\eqn\super{
W = \tr X^i X^j Y^{[ij]} + \ldots .  } 
Such a term in the superpotential is also natural from the CFT point
of view and as discussed in \bdlr, it can be computed in the topologically
twisted model.  The non-zero amplitudes are those in which the
operators combine to saturate the fermion zero modes; in terms of
the translation to forms on $\BC^5$ given above they are the amplitudes in
which the product of the forms involved produces a top form on $\BC^5$,
which produce exactly \super.  

\subsec{Quiver gauge theory}

We now apply the orbifold projection to derive the world-volume theory
of boundary states on $\BC^5/\Gamma$.  We will discuss
$\Gamma\cong\BZ_K$ in detail here, though similar
considerations would apply to nonabelian groups, as for $\BC^2/\Gamma$
in \johnsonmyers.  It will be a quiver theory with $K$ nodes, chiral
superfields in the ${\bf 5}$ of $SU(5)$ $X^i_{M,M+w_i}$, 
chiral superfields
in the $\overline{\bf 10}$ 
$Y^{ij}_{M,M-w_i-w_j} \equiv -Y^{ji}_{M,M-w_i-w_j}$,
and the restriction of \super,
\eqn\gepnerW{
W = \sum_{M,i,j} X_{M,M+w_i}^{i} X_{M+w_i,M+w_i+w_j}^{j} Y_{M+w_i+w_j,M}^{ij}.
}

Indeed all of this data agrees with the explicit CFT results of
\refs{\rs,\bdlr}.  B boundary states in these models are characterized
by labels $L_i$ in each minimal model factor and $M=\sum w_j M_j$ in
$[0,2K-1]$.  In particular, if we define the states $\ket{M}$ with a
given $M$ and all $L_i=0$, and the operator $g$ acting as
$M\rightarrow M+2$, then we can write the intersection matrix between
the $L=0$ states as
\eqn\intmatrix{\eqalign{
I_G &= \prod_j (1-g^{w_j})  \cr
&= \sum_{k_j=0,1} (-)^{\sum k_j} g^{\sum k_j w_j}
}}
which agrees with the massless Ramond spectrum we described.  This term
in the superpotential can be checked from CFT \ilka.

In general the superpotential will contain higher order terms as well.
These are computable in CFT and are also topological, but not too much is
known about them at present.  Our results so far are consistent with the idea
that in the theories in which the low energy description is justified
(we will explain this point shortly), such terms are absent, but this remains
to be seen.

The final item required to complete the specification of the
world-volume gauge theories is the Fayet-Iliopoulos terms for the
$U(1)^K$ subgroup of the gauge group.  As pointed out in
\refs{\pistable,\category}, these can be determined from the masses
of the bosonic superpartners of the massless fermions, which are also
known from CFT.  These superpartners can be defined using the spectral
flow operator on either brane (they produce the same result up to a
phase) and in the Gepner models under discussion are obtained by
multiplying the top chiral primaries in a subset of the individual minimal
model factors.  The result is that a boson on a link from $M$ to $M+w$ 
has $m^2 = -w \lambda$ where we consider $X^i_{M,M+w_i}$ 
as ``forward'' links (so these are tachyonic) and $Y^{[ij]}_{M,M-w_i-w_j}$
as ``backward'' links (so these are massive).
$\lambda$ is computable and order string scale.

The FI terms must then reproduce these masses
$$
m^2_{ij} = \zeta_i - \zeta_j
$$ 
where $\zeta_i$
is the FI term for the $U(1)$ of the $i$'th brane.
In fact one can argue directly for this structure from general properties
of $\CN=2$ CFT: the $\zeta_i$ and thus $m^2$ are directly related to the
overall $U(1)$ charge and thus the phases of the central charges 
of the two branes \category.

This determines the FI terms to be $\zeta_k=k \lambda$, but one immediately
notices that this cannot reproduces all the masses in all theories.
The condition for it to work is that we do not have links going 
``all the away around the clock,'' for example closed loops with only
$X$ fields.  It does not forbid closed loops involving both $X$ and $Y$.

This is a condition on the allowed fractional
brane content: if all types of fractional branes are present it will
fail, but it can be satisfied by excluding some types of fractional branes.
If it fails, the masses cannot all be reproduced
in this low energy field theory description, which probably signals its
breakdown.  As we will see, such configurations typically restrict
to branes on the Calabi-Yau with zero RR charge (or with simpler
realizations) and would thus be expected to decay to the vacuum (or the
simpler realization); this process would then not have a description
purely in the low energy theory.

In conclusion, we have derived the low energy theories of combinations
of the $L=0$ Gepner model branes by adapting the orbifold construction
to $\BC^5/\BZ_K$ in a way suggested by linear sigma model
considerations.  The result is a quiver gauge theory very analogous to
those for $\BC^3/\Gamma$ orbifolds.  The construction generalizes 
straightforwardly to general quotients $\BC^r/\Gamma$.

\subsec{Bound states}

We now make the general claim that supersymmetric vacua of these
theories with unbroken gauge symmetry $U(1)$ correspond to general
(classical) bound states of these rational branes.  If one keeps all
the modes of the open string theory (e.g.\ by using string field
theory), this claim seems difficult to dispute.  A less obvious
claim is that many bound states can be described purely within
the theory obtained by keeping the chiral primaries, in other words
the theories we just derived.  The potential
problem is that the FI terms and thus the
vevs of the fields at the supersymmetric vacuum have string-scale
values.

Nevertheless, as we discussed, in a large subset of the theories we
discussed, those with chiral fields whose masses can be reproduced by
FI terms, there is no {\it a priori}\/ reason for this description to break down.
A basic test of it which can be done is to construct
non-rigid ($L>0$) rational branes as bound states of the $L=0$ branes and
check that the dimension of the moduli space comes out right.
Some non-trivial examples of this for the quintic were
discussed in \noncompact, and further examples will be discussed in
\inprogress. 

Now, work on non-BPS brane configurations in flat space and other
simpler examples does support the claim that in many cases a good
qualitative description can be obtained just using tachyons and
massless fields. Thus it should probably not be too surprising that
this works in some subset of the theories we just derived.  Whether
this works in all the theories which {\it a priori}\/ appear sensible, and
whether string field theory or similar frameworks can provide a
more general description, are important questions for further work.

\newsec{The large volume interpretation of the fractional branes}

In \noncompact, it was noticed that the large volume interpretation of the
fractional branes of the $\BC^3/\BZ_3$ orbifold found using mirror symmetry 
\dg\ gave the same bundles which form the natural basis (due to Beilinson)
for the general
construction of bundles on $\P^2$, which is the exceptional divisor of the
resolution of $\BC^3/\BZ_3$.

As it turns out, recent mathematical work has led to a very general
conjecture for the higher dimensional analog of this correspondence,
which we will be able to apply directly to our $\BC^5/\Gamma$ theories
\refs{\reid,\bkr,\itonakajima}.

The idea is to generalize the famous McKay correspondence
\mckay\ between discrete subgroups $\Gamma$ of $SU(2)$ and finite simply laced
Dynkin diagrams to discrete subgroups $\Gamma\subset SU(N)$.  More
specifically, one has a relation between the representation ring of
$\Gamma$ (which is directly encoded in the Dynkin diagram) and a basis
for the exceptional cycles in a resolution of a $\BC^2/\Gamma$
singularity.  According to \reid, the idea that a higher dimensional
generalization should exist actually has its origins in the study of
orbifolds in string theory, and the observation that the Euler number
of a resolved $\BC^3/\Gamma$ singularity equals the number of
conjugacy classes of $\Gamma$.

The precise version of this idea which has been generalized is a duality
between the category of sheaves on $X\cong\BC^n/\Gamma$,
and the category of sheaves with compact support,
which for $X\cong\BC^n/\Gamma$ will be sheaves supported
at the origin (or, if a partial resolution has been performed,
on the exceptional divisor).  

In making this precise, one must work with specific categories.
The duality can be made quite concrete, as was done by Ito and Nakajima in
\itonakajima, where the dual objects are constructed as explicit complexes of
line bundles.  This should allow making a detailed identification between
quiver representations and large volume sheaves along the lines of
\noncompact.  In \bkr, the duality was shown to be an equivalence between
derived categories, which should allow proving the analog of Beilinson's
theorem for this case.  Although less concrete, this is still a very strong
statement about the relation between the two categories.

Here we will content ourselves with deriving the K-theory classes which
correspond to the
dual basis.  We will then restrict these to the Calabi--Yau and compare with
the predictions of mirror symmetry in a solved example.

The mathematics of the generalized McKay correspondence is clearly described
in \refs{\reid,\bkr,\itonakajima}\ and thus in the remainder of this section
we concentrate on describing the ideas for physicists.

\subsec{Orbifold resolution, tautological line bundles, and dual bases}

The problem of resolving singularities $X=\BC^2/\Gamma$ has a long
mathematical history, going back to Klein (see \reidbourbaki\ for some
of this background).  Such a singular variety $X$ can be resolved
to a smooth space $Y$ with non-trivial $H^2(Y)$ and intersection form
given by the Cartan matrix of the extended ADE Dynkin diagram
associated to the subgroup $\Gamma\in SU(2)$.

The most basic string theory application of this is to the duality
between \IIa\ strings on K3 and heterotic strings on $T^4$ \hulltown,
where \IIa\ D$2$-branes wrapped on the resolved (or ``exceptional'')
cycles provide the nonabelian gauge bosons 
of the enhanced ADE gauge symmetry predicted by duality.

The most direct connection between this geometry and the structure of
the group $\Gamma$ appears in the McKay correspondence.  The McKay
quiver associated to $\Gamma$ has a node for every irrep $r_i$ of $\Gamma$,
and a link from $r_i$ to $r_j$ for every component $r_j$ in
$r_{def} \otimes r_i$, where $r_{def}$ is the representation by which
$\Gamma$ acts on $\BC^2$.  The result is a quiver which can be simply
obtained from the ADE extended Dynkin diagram by replacing each link
of the latter by a pair of links of opposite orientation.

This construction can also form the basis of an explicit construction
of the resolved space, as was done by Kronheimer \refs{\kronone,\krontwo}.  
Physically
the same construction appears in defining D-branes on the quotient
space, and provides an explicit gauge theory description of the
resolution \dm.  It furthermore provides an explicit description of
the branes wrapped on the exceptional cycles; these are ``fractional
branes'' obtained by using irreducible representations in the quotient
construction.  One thus has the basic prediction that there should
exist a natural basis for $H^2(Y)$, or better the K-theory of the
resolved singularity, labeled by irreducible representations of $\Gamma$.

In this context the relation between the Dynkin diagram and the
intersection form has a physical interpretation as well:
each link corresponds to a hypermultiplet coming from open strings
stretched between a pair of fractional branes; their number can be
computed from the index theorem and is equal to the intersection
number.

An even more general physical system was studied in \dm, containing both 
D$p$-branes at points in $X$ and D$p+4$-branes extending in $X$.
It was found to reproduce a construction of general self-dual
gauge fields on the resolved singularity due to Kronheimer and Nakajima 
\kronnak.
Now both types of branes are labeled by a choice of group representation,
and each can be associated to a quiver node.  Let $R_i$ be the D$p+4$
node corresponding to $r_i$ and $S^j$ be the D$p$ node corresponding to $r_j$.
The spectrum of $(p,p+4)$-strings between a pair $(R_i,S^j)$
is also determined by the orbifold 
projection and one finds the number of hypermultiplets to be $\delta_i^j$.
As in our previous discussion, this implies that the intersection form
between the two types of branes should be
\eqn\dualrelation{
\vev{R_i,S^j} = \delta_i^j.
}

Now the interpretation of the D$p+4$ (extended) branes as bundles is
rather clear, at least far from the singularity.  The orbifold projection
acts on the Yang--Mills connection as
\eqn\orbgauge{
\gamma A_i(z) \gamma^{-1} = r_i^j A_j(g(z)).
}
This tells us that scalar matter in the fundamental, i.e., a section of the
associated bundle, must transform as
\eqn\orbmatter{
\gamma \phi(z) = \phi(g(z)).
} 
A particularly simple case is to take $\gamma$ to be the regular
representation, in which case we can consider $\phi(z)$ as a
vector-valued field indexed by an element of $\Gamma$, so \orbmatter\ becomes
\eqn\orbtaut{
\phi_{gh}(z) = \phi_h(g(z)).
} 
This bundle is referred to
as the ``tautological bundle'' over the quotient space.  It can be
decomposed as a direct sum over bundles $R_i$ associated to irreps $\gamma$
which if $\Gamma$ is abelian are line bundles; these are the
tautological line bundles.  

The dual relation \dualrelation\ then determines the bundles $S_j$.
On a noncompact space $X$, the natural duality for K-theory (just as
for cohomology) is between $K(X)$ and the K-theory of bundles with
compact support $K_c(X)$, meaning bundles over compact submanifolds of
$X$.  Thus the bundles $S^j$ naturally live in $K_c(X)$ and provide a
preferred basis for it.  These are the bundles associated to the
fractional D$p$-branes.  

Given the intersection form in an explicit basis, we can make this
definition quite concrete.  For example, if we have
\eqn\Rintform{
\vev{R_i,R_j} \equiv (I^{-1})_{ij},
}
then we can write
\eqn\defS{
S^j = I^{ij} R_i 
}
for which
\eqn\Sintform{
\vev{S^j,S^k} = I^{jk} .
}
As in \itonakajima,
the relation \defS\ can be used to define the $S^j$ as complexes built
from the bundles $R_i$.  In terms of the K-theory classes, \defS\ becomes
\eqn\KS{
[S^j] = I^{ij} [R_i] ,
}
a simple explicit formula for the K-theory classes of the fractional
branes given those of the tautological line bundles.  

Restricting these bundles or their
classes to a subvariety, such as a Calabi--Yau embedded in the exceptional
divisor, is a standard operation:
let $V^j = S^j|_{CY}$ be these restrictions.
Thus we can define an intersection
form on the Calabi--Yau 
\eqn\intCY{
I_{CY}^{jk} = 
\vev{V^j,V^k}_{CY} = \int_{CY} \ch({\overline{V}}^j) \ch(V^k) 
\td(CY),}
and the conjecture is that
\eqn\intconj{
I_{CY} = I_G
}
where $I_G$ is the intersection form
\intmatrix\ of section 2.

The result is a physically motivated
prediction for the K-theory classes of the rational B boundary states,
which we will test against results derived using mirror symmetry.
Indeed, the example of the quintic discussed in \noncompact\ is already
a non-trivial test, as the procedure we just described leads to Beilinson's
dual bases in the case of $\BC^n/\BZ_n$, which as checked there agree
with the results of \bdlr.

Although \KS\ is the formula we will test in this paper,
let us emphasize that \defS\ provides a definition of
the fractional branes $S^j$ as holomorphic objects, not just K-theory
classes.  This is made quite explicit in \refs{\bkr,\itonakajima},
where the dual bases in \dualrelation\ are used to construct a resolution
of the diagonal, which can be used to prove Beilinson's theorem for these
spaces.  This leads to explicit large volume interpretations of general
bound states of the fractional branes as complexes of sheaves, as we will
discuss in future work \inprogress.

So far, none of our definitions had any real dependence on the
dimension of $X$; we could make the same discussion for $\BC^n/\Gamma$
for any $n$.  The point where such dependence will come in is when we
discuss the resolution of the singular space $X$ in detail.  Indeed,
unless we can resolve $X$, it is not obvious in what sense the $R_i$
can be thought of as bundles or how to compute their K-theory classes.
General theory \fulton\ does tell us that given a resolution $Y$ of $X$,
there will be a natural lift of these K-theory classes to $Y$, but 
we might expect this to depend on the particular resolution we choose.

Thus we need to discuss the resolution of $X$ in more detail.  One
idea which has been used with great success in the math literature has
been to use subspaces of the Hilbert scheme of $N=|\Gamma|$ points on
$\BC^n$ which are invariant under $\Gamma$.  It has been shown for
$n=2$ and all $\Gamma$, and $n=3$ and abelian $\Gamma$, that such a
subspace provides a canonical complete resolution $Y$ of $X$.  The
definition of tautological bundle then lifts naturally to $Y$,
and the story can be completed in this framework.

For $n>3$ there are known examples in which this construction does not
produce a complete resolution.  Moreover, the Hilbert scheme becomes
progressively more difficult to work with in higher dimensions.

An alternate approach is to define the quotient $X$ as the moduli
space of a quiver gauge theory, and then find the resolution $Y$ by
the usual procedure of turning on Fayet--Iliopoulos terms.  This
approach was successfully used for $\BC^3/\Gamma$ by Ito and Nakajima
and is clearly well motivated in our D-brane application, so we shall
follow it below.  One disadvantage of this approach is that the choice
of Fayet--Iliopoulos terms generally translates into a choice of
resolution; it is not obvious that any of these is preferred.
However, as we argued, the Gepner models produce quiver gauge theories
come with a natural choice of FI terms, so we should try to make the
construction work with these.

\newsec{Orbifolds via toric methods}

A general procedure for analyzing abelian orbifolds as toric varieties
was given in \dgm.  We will briefly review this, and give the definition
of the tautological line bundles in this context.

D-branes in orbifold backgrounds are described by supersymmetric
world-volume gauge theories, as was shown in \dm\ for orbifolds in
flat space and as we have argued here for Landau--Ginzburg orbifolds.
The resolved orbifold will be the moduli space of supersymmetric vacua
of the regular representation theory.

In physical terms, a toric variety can be defined as the moduli space
of vacua for an abelian $\CN=1$ supersymmetric gauge theory with no
superpotential.  While the moduli space of vacua for a general
supersymmetric gauge theory does not admit a toric realization,
theories for which the F-flatness constraints can be written as
relations between monomials do.  

In the general class of theories we described,
the F-flatness conditions are indeed relations between monomials:
they are
\eqn\Fflatness{
X^i_{M,M+w_i}X^j_{M+w_i,M+w_i+w_j}=X^j_{M,M+w_j}X^i_{M+w_j+w_i},}
where $M=1,\ldots,|\Gamma|$ labels the nodes of the quiver diagram 
and $X^i_{M,M+w_i}$ are the chiral multiplets. \foot{
We are only considering the special case $Y=0$ here, as this is what
makes direct contact with \itonakajima\ and the resolution to weighted
projective spaces.  The moduli $Y$ appear to be connected with
deformations of bundles which appear after restriction to the CY \inprogress.}
As explained in \dgm,
the solutions to these constraints are parameterized by an 
affine toric variety ${\cal Z}\subset \BC^{d|\Gamma|}$, which has been 
called {\it the variety of commuting matrices}\/ in \partres.

The idea which allows describing this as a toric variety
can be illustrated with the variety $X$ defined by the simple
relation $xy=wz$.  Let us solve for $z$ as $z=xy/w$.  We can then
describe the space of functions on the variety, as the functions
$f(w,x,y)$ which are generated by multiplying the monomials $w$, $x$
and $y$ and the monomial $xy/w$.  In other words, the presence of $z$
is described by admitting more functions than we would on $\BC^3$.
The set of exponents of these monomials is the cone $M_+$ generated by
positive integral
combinations of the vectors $(1\ 0\ 0)$, $(0\ 1\ 0)$, $(0\ 0\ 1)$
and $(-1\ 1\ 1)$.

The same data can be described by giving the dual cone $N_+$ of vectors
satisfying $n\cdot m\ge 0$.  In this case it would be generated by 
$n_a=(1\ 1\ 0)$, $n_b=(1\ 0\ 1)$, $n_c=(0\ 1\ 0)$ and $n_d=(0\ 0\ 1)$.

Now we can describe the space of functions on $X$ by associating variables
with these generators of $N_+$, say $a$, $b$, $c$ and $d$, and writing
monomials in these variables.  The non-trivial data about $X$ is now
expressed in the relations between the generators.  In our example there
is a single relation, $n_a+n_d=n_b+n_c$.

The important fact is now that
{\bf constraints are dual to gauge invariances},
where duality is in the linear algebra sense.
This is fairly obvious on reflection but 
can be best seen by using the language of exact sequences.
Consider a sequence
$$0\mapr A \mapr^f B \mapr^g C \mapr 0.$$
Its exactness means that $g\cdot f=0$.

One interpretation we could make of this is that $g=0$ expresses
a set of constraints on the space $B$, parameterized by elements of $C$.
The map $f$ would then be an explicit set of solutions to the constraints,
parameterized by elements of $A$.

Another possible interpretation is that we have abelian
gauge symmetries acting on the space $B$, described by the image of
the map $f$ and parameterized by elements of $A$.  We could then regard
$C$ as the gauge invariant subspace or quotient $B/A$.  This formulation
is the best when we can use it, as it avoids the need to make an explicit
choice of a gauge slice; if we needed to exhibit a slice in $B$ we would
need to choose a partial inverse $h$ of $g$ satisfying $g\cdot h=1|_C$;
$h$ would then give a map from $C$ to the slice.

The point now is that duality (in the linear algebra sense) reverses
all the arrows and thus the roles of the maps $f$ and $g$.  This leads
to duality between constraints and gauge invariances.

In this example,
the dual relation between $M_+$ and $N_+$
implies that constraints on the $M$ monomials will lead to gauge invariances
for the $N$ monomials.  This is formalized by writing the space $X$ as the
spectrum of the algebra of monomials.  Defining this algebra as
$\Hom(M,\BC)$ allows applying the previous discussion; see for example \cox.

Thus, the relation $n_a+n_d=n_b+n_c$ of our example should translate
into a gauge invariance, with $U(1)$ acting on
the four variables $(a\ b\ c\ d)$ with the charges $(1\ -1\ -1\ 1)$.
We can test this claim by writing out the gauge invariant
monomials and checking that they satisfy the relation.
Indeed, these are $ab$, $ac$, $db$ and $dc$, which can be
identified with the original $w$, $x$, $y$, and $z$ satisfying the
relation $xy=wz$.

This type of realization, in which the relations between generators of
$N_+$ are interpreted as gauge invariances, is completely general, and
gives a method for turning the F-flatness constraints into abelian
gauge invariances.  Thus we can realize the final moduli space of
vacua entirely as a toric variety.  The main difference between the
original abelian gauge invariances and the newly generated ones is
that the former will typically come with FI terms, while the latter
will not.

$U(1)$ gauge groups with non-zero FI terms will typically be associated
to non-trivial topology in the quotient, as is also clear in our
example: turning on the FI term produces a bundle over $\P^1$ (the
``small resolution'' of the conifold).  We will make a related 
precise statement in the next subsection, which we will use to justify
associating the tautological line bundles with $U(1)$'s coming with
non-zero FI terms and thus with nodes in the original quiver.

\subsec{A concise presentation}

We describe the variety of commuting matrices ${\cal Z}$ in
terms of its lattice of monomials $M$, and the positive cone
$M_+$ within that lattice.  The lattice $M$ is given as the quotient 
\eqn\monlat{
0\rightarrowbox{12pt}{}R\rightarrowbox{12pt}{{\widehat R}^t}
 \BZ^{d|\Gamma|}\rightarrowbox{12pt}{K^t}M
\rightarrowbox{12pt}{} 0}
where $R$ is the lattice of relations corresponding to the 
F-flatness constraints.  It can  be easily checked that 
\eqn\ranks{
\hbox{rk}(R)=(d-1)(|\Gamma|-1),\qquad \hbox{rk}(M)=d+|\Gamma|-1.}
The cone of monomials $M_+$ consists of all elements of $M$ 
with nonnegative components 
$M_+=M\cap\BR^{\dgam}_+$.
The linear map ${\widehat R}$ in \monlat\ can be represented by a
$(d-1)(|\Gamma|-1)\times \dgam$ matrix whose rows correspond to 
the relations \Fflatness. $K$ is a $\dgam\times (d+|\Gamma|-1)$ 
matrix whose columns form an integral basis for the kernel of 
${\widehat R}$. The column vectors of $K$ form an integral basis
of $M$, which is therefore isomorphic to $\BZ^{d+|\Gamma|-1}$. 

As explained in \refs{\dgm,\partres} and just above,
the variety ${\cal Z}$ can be 
alternatively represented as a holomorphic quotient 
$\BC^c/(\BC^*)^{c-d-|\Gamma|+1}$. Here $c$ is the number of generators 
of the dual cone $N_+\subset N=\hbox{Hom}(M,\BZ)$. In this case, 
the generators of $N_+$ correspond to $c$ homogeneous variables 
and we have $d+|\Gamma|-1$ relations generating the charge lattice. 
This data can be conveniently summarized in an exact sequence 
\eqn\tordata{
0\rightarrowbox{12pt}{} S\rightarrowbox{12pt}{Q^t}
\BZ^c\rightarrowbox{12pt}{T}N\rightarrowbox{12pt}{}0}
where $S$ is the lattice of charges. $T$ is a $(d+|\Gamma|-1)\times c$ 
matrix whose column vectors generate $N_+$. $Q$ is the transpose of 
the kernel of $T$. 

At the next stage, we have to solve for D-flatness constraints, 
which is equivalent to taking a symplectic quotient of ${\cal Z}$
by the effective gauge group $G$ of the quiver. Note that the diagonal 
$U(1)$ is ungauged, therefore we have $G=U(1)^{|\Gamma|-1}$. 
The action of $G$ on the monomials can be represented as a linear 
map $\Delta: \BZ^{\dgam}\rightarrow \BZ^{|\Gamma|-1}$ assigning to 
each monomial a vector of charges. Since the relations \Fflatness\
are invariant under $G$, this map factors through 
$p:\BZ^{\dgam}\rightarrow M$, i.e., there exists $V:M\rightarrow 
\BZ^{|\Gamma|-1}$ such that $\Delta = V\circ p$. 
By dualizing, one obtains the maps 
$\Delta^t:\left(\BZ^{|\Gamma|-1}\right)^*\rightarrow 
\left(\BZ^{\dgam}\right)^*$ and $V^t:\left(\BZ^{d|\Gamma|-1}\right)^*
\rightarrow N$ which fit in the following diagram 
\diagram[Postscript=dvips]
&   &         & 0 &         &                          &         & 
&  & \\
& & & \uTo_{} & & & & & & \\
& 0 & \rTo_{} & N & \rTo^{} & \left(\BZ^{\dgam}\right)^* & \rTo^{} & 
R^* & \rTo^{} & 0 \\
&   &         & \uTo^T\dTo_{U^t} & \luTo^{V^t} & \uTo_{\Delta^t} & 
& & & \\
& & & {\BZ}^c &\lTo^{U^tV^t} & \left(\BZ^{|\Gamma|-1}\right)^* & & 
& & \\
& & & \uTo^{Q^t} & & & & & & \\
& & & S & & & & & & \\
& & & \uTo_{} & & & & & & \\
& & & 0 & & & & & & \\
\enddiagram
The map $U^t$ is a quasi-inverse of $T$ satisfying 
$TU^t=I_{d+|\gamma|-1}$ which is required in order 
to express the action of $G$ on ${\cal Z}$. More precisely, 
the final moduli space is represented by the toric data 
\eqn\modsp{
0\rightarrowbox{12pt}{} S\oplus\left(\BZ^{|\Gamma|-1}\right)^*
\rightarrowbox{12pt}{(Q^t,(VU)^t)}{\BZ}^c
\rightarrowbox{12pt}{} {\widetilde N}\rightarrowbox{12pt}{} 0}
where ${\widetilde N}$ is a rank $d$ lattice. 
Note that the matrix of charges ${\widetilde Q}$ is obtained by 
concatenating $Q$ and $VU$
\eqn\chmatrix{
{\widetilde Q}=\left[\matrix{Q\cr VU\cr}\right].}
To summarize, note that we have interpreted the $U(1)$ gauge groups 
of the quiver theory as a subset of the generators of the charge 
lattice of the toric moduli space $X$. 
It is known that each such generator corresponds to a Weil divisor class 
on $X$ (\fulton\ 3.4 or \cox\ section 2); the basic idea is that 
\tordata\ is dual to the exact sequence expressing divisor classes
as divisors (rays in $N_+$) modulo functions (elements of $M$).

Since these $U(1)$ gauge groups are labeled by a choice of $\Gamma$
irrep, this construction produces a canonical divisor class
for each irrep.  This will be our definition of
the tautological line bundles $R_k$. 

The question of whether these divisors actually correspond to line bundles
(live in $\Pic(X)$) {\it a priori}\/
depends on the particular FI terms and subdivision
of the toric fan which we take.  
We will check this explicitly for our example;
a more general treatment might use the methods of \partres.

\subsec{Limitations of the method}

As we discussed in section 3, the identification of the dual bases
with actual K theory classes implicitly requires that the space $X$ be
completely resolved.  It has not been proven that the procedure of
\dgm\ will always produce a complete resolution, even if one exists.

Even worse, in dimensions four and above, not all $SU(n)$ orbifold
singularities are crepant (admit Calabi-Yau resolutions), and it is
not clear how to define the generalized McKay correspondence in these
cases.  This is a problem as many Gepner models construct Calabi-Yaus
as hypersurfaces in weighted projective spaces containing such
singularities.  If the hypersurface avoids the singularity, one has a
completely sensible Calabi-Yau and geometrical interpretation, but the
method presented here does not immediately apply.  It is an important
open problem to find a method which handles these cases.

There are also many cases for which this is not a problem, and we proceed
to test the method for one of them.

\newsec{An example: $\WP^{1,1,2,2,2}$}

In the previous section we have proposed a general toric algorithm 
for determining the tautological line bundles $R_k$ on a (resolved)
weighted projective space. We have also conjectured that the resulting
line bundles
$R_k$ are K-theory generators dual to the classes $S_k$ corresponding 
to the fractional branes. Here we test this conjecture for 
the two parameter model $\WP^{1,1,2,2,2}$ discussed in \twopar.
The B boundary states for this model have been considered in 
\lerche.

Before presenting the details of the computation, some background 
on the geometry of $\WP^{1,1,2,2,2}$ may be helpful. This is a singular 
weighted projective space whose toric resolution is defined 
by the following data \quantcoh\
\eqn\toricA{
\matrix{& & x_0 & x_1 & x_2 & x_3 & x_4 & x_5 \cr
&g_1(\lambda)& 1 & 1 & 1 & 1 & 0 & 0 \cr
&g_2(\lambda)& 0 & 0 & 0 &-2 & 1 & 1 \cr}}
with disallowed set 
\eqn\toricB{
F=\{x_0=x_1=x_2=x_3=0\}\cup\{x_4=x_5=0\}.}
The resulting smooth toric variety will be denoted by $W$ 
for simplicity. 
The Picard group $\hbox{Pic}(W)$ is generated by two divisors 
$H, L$ with relations
\eqn\relA{
H^3(H-2L)=0,\qquad L^2=0}
and intersection numbers 
\eqn\intA{
H^4=2,\qquad H^3L=1,\quad H^2L^2 = HL^3 = L^4 = 0.}
The Calabi--Yau  hypersurface $M\subset W$ is the zero locus of a 
generic section of the anticanonical line bundle $-K_W=\CO(4H)$. 

Certain details on the geometry of $M$ will also be needed 
in the following \twopar. For simplicity, let $H,L$ also denote 
the restrictions of the divisor classes to $M$. The meaning will 
be clear from the context. Then we have the intersection numbers 
\eqn\intB{
(H^3)_M=8,\qquad (H^2L)_M=4,\qquad (HL^2)_M= (L^3)_M=0.}
The cone of curves on $M$ is generated by $(h,l)$ \twopar\ 
such that 
\eqn\intC{\eqalign{
& (H\cdot h)_M = 1 ,\qquad (H \cdot l)_M =0\cr
& (L\cdot h)_M = 0,\qquad (L\cdot l)_M =1.\cr}}
Moreover, we have the intersection relations 
\eqn\intD{\eqalign{
 4l=(H^2-2HL)_M\ &\Rightarrow\  H(H^2-2HL)=l\cr
4h=(HL)_M\ &\Rightarrow\  H^2L=h.\cr}}
Finally, the second Chern class of $M$ is 
\eqn\chernA{
c_2(M)= 56h + 24l.}

\subsec{Tautological line bundles}

We start by determining the quiver moduli space and the tautological 
line bundles for $\BC^5/\Gamma$, where $\Gamma=\BZ_8$ acts on $\BC^5$ 
as 
\eqn\actA{
(Z^1,Z^2,Z^3,Z^4,Z^5)\rightarrow (\omega Z^1,\omega Z^2,\omega^2 Z^3,
\omega^2 Z^4,\omega^2 Z^5),\qquad \omega=e^{2\pi i\over 8}.}
As discussed in section 2.3, the associated quiver theory has eight 
nodes labeled by $M=0,\ldots, 7$. For each node, we have five chiral 
multiplets $X^{1,2}_{M,M+1},\ X^{3,4,5}_{M,M+2}$. 
The superpotential \super\ yields eighty F-flatness conditions 
\Fflatness,\ out of which only twenty-eight are independent.
We first determine the moduli space of the quiver gauge theory
using toric methods as explained in the previous section. 
Then we find similarly the tautological line bundles $R_k$.

The equations \Fflatness\ can be solved in terms of the twelve 
independent variables 
$X^1_{01},X^2_{01},X^3_{02},X^4_{02},X^5_{05},X^1_{12},\ldots,
X^1_{70}$
\eqn\Fsol{\eqalign{
& X^2_{M,M+1}={X^2_{0,1}\over X^1_{0,1}}X^1_{M,M+1},
\qquad M =1,\ldots, 7\cr
& X^i_{M,M+2}={X^i_{02}\over X^1_{01}}
{X^1_{M,M+1}X^1_{M+1,M+2}\over X^1_{12}},\qquad i=3,4,5.,\
M=1,\ldots,7.\cr}}
We obtain therefore twenty-eight vectors in $\BR^{12}$ spanning 
the cone of monomials $M_+$. The dual cone $N_+$ is spanned 
by twenty-one twelve dimensional vectors, whose coordinates form a
$12\times 21$ matrix $T$ 
\eqn\Tmatrix{
T=\left[\matrix{
 0 & 0 & 0 & 0 & 0 & 0 & 0 & 0 & 0 & 0 & 0 & 0 & 0 & 0 & 0 
& 1 & 1 & 1 & 1 & 1 & 1 \cr
 0 & 0 & 0 & 1 & 0 & 0 & 0 & 0 & 0 & 0 & 0 & 0 & 0 & 0 & 0 
& 1 & 1 & 1 & 1 & 1 & 0\cr
 0 & 0 & 1 & 0 & 0 & 0 & 0 & 0 & 0 & 0 & 1 & 0 & 0 & 0 & 0 
& 1 & 0 & 0 & 0 & 0 & 0\cr
 0 & 1 & 0 & 0 & 0 & 0 & 0 & 0 & 0 & 0 & 1 & 0 & 0 & 0 & 0 
& 1 & 0 & 0 & 0 & 0 & 0\cr
 1 & 0 & 0 & 0 & 0 & 0 & 0 & 0 & 0 & 0 & 1 & 0 & 0 & 0 & 0 
& 1 & 0 & 0 & 0 & 0 & 0\cr
 0 & 0 & 0 & 0 & 0 & 0 & 0 & 0 & 0 & 0 & 1 & 1 & 1 & 1 & 1 
& 0 & 0 & 0 & 0 & 0 & 1\cr
 0 & 0 & 0 & 0 & 0 & 0 & 0 & 0 & 0 & 1 & 0 & 0 & 0 & 1 & 1 
& 0 & 1 & 1 & 1 & 1 & 1\cr
 0 & 0 & 0 & 0 & 0 & 0 & 0 & 0 & 1 & 0 & 0 & 1 & 1 & 0 & 0
& 0 & 0 & 0 & 1 & 1 & 1\cr
 0 & 0 & 0 & 0 & 0 & 0 & 0 & 1 & 0 & 0 & 0 & 0 & 1 & 1 & 1
& 0 & 1 & 1 & 0 & 0 & 1\cr
 0 & 0 & 0 & 0 & 0 & 0 & 1 & 0 & 0 & 0 & 0 & 1 & 0 & 0 & 1
& 0 & 0 & 1 & 1 & 0 & 1\cr
 0 & 0 & 0 & 0 & 0 & 1 & 0 & 0 & 0 & 0 & 0 & 0 & 1 & 1 & 0
& 0 & 1 & 0 & 0 & 1 & 1\cr
 0 & 0 & 0 & 0 & 1 & 0 & 0 & 0 & 0 & 0 & 0 & 1 & 1 & 1 & 1
& 0 & 0 & 1 & 1 & 1 & 1\cr
}\right].}
The columns of $T$ correspond to homogeneous coordinates 
of the variety of commuting matrices ${\cal Z}$. The transpose of the 
kernel of $T$ determines a charge matrix $Q$
\eqn\Qmatrix{\eqalign{
&Q=\cr
&\left[\matrix{
1 & 1 & 1 & 0 & -1& -1& 0 & 0 & -1& -1& 0 & 0 & 0 & 0 & 0 
& -1& 0 & 0 & 0 & 1 & 0\cr
1 & 1 & 1 & 0 & -1& 0 & -1& 0 & -1& 0 & -1& 1 & 0 & 0 & 0 
& 0 & 0 & 0 & 0 & 0 & 0\cr
1 & 1 & 1 & 0 & -1& -1& 0 & -1& 0 & -1& -1& 0 & 0 & 1 & 0 
& 0 & 0 & 0 & 0 & 0 & 0\cr
2 & 2 & 2 & 1 & -1& -1& -1& -1& -1& -1& -1& 0 & 0 & 0 & 0 
& -1& 0 & 0 & 0 & 0 & 1\cr
1 & 1 & 1 & 0 & -1& 0 & -1& -1& 0 & -1& -1& 0 & 0 & 0 & 1
& 0 & 0 & 0 & 0 & 0 & 0\cr
1 & 1 & 1 & 0 & -1& -1& 0 & -1& -1& 0 & -1& 0 & 1 & 0 & 0
& 0 & 0 & 0 & 0 & 0 & 0\cr
1 & 1 & 1 & 0 & 0 & -1& 0 & -1& 0 & -1& 0 & 0 & 0 & 0 & 0
& -1& 1 & 0 & 0 & 0 & 0\cr
1 & 1 & 1 & 0 & -1& 0 & -1& -1& 0 & -1& 0 & 0 & 0 & 0 & 0
& -1& 0 & 1 & 0 & 0 & 0\cr
1 & 1 & 1 & 0 & -1& 0 & -1& 0 & -1& -1& 0 & 0 & 0 & 0 & 0
& -1& 0 & 0 & 1 & 0 & 0\cr
}\right].\cr}}
In order to find the action of the quiver $U(1)$ gauge groups
on ${\cal Z}$, we have to 
choose a $12\times 21$ matrix $U$ such that 
\eqn\Udef{
TU^t=I_{12}.}
We pick $U$ of the form 
\eqn\Umatrix{\eqalign{
&U=\cr
&\left[\matrix{
 0 & 0 & 0 &-1 & 0 &-1 & 0 &-1 & 0 & -1 & 0 & 0 
& 0 & 0 & 0 & 0 & 1 & 0 & 0 & 0 & 0 \cr
 0 & 0 & 0 & 1 & 0 & 0 & 0 & 0 & 0 & 0 & 0 & 0 
& 0 & 0 & 0 & 0 & 0 & 0 & 0 & 0 & 0 \cr
 0 & 0 & 1 & 0 & 0 & 0 & 0 & 0 & 0 & 0 & 0 & 0 
& 0 & 0 & 0 & 0 & 0 & 0 & 0 & 0 & 0 \cr
 0 & 1 & 0 & 0 & 0 & 0 & 0 & 0 & 0 & 0 & 0 & 0 
& 0 & 0 & 0 & 0 & 0 & 0 & 0 & 0 & 0\cr
 1 & 0 & 0 & 0 & 0 & 0 & 0 & 0 & 0 & 0 & 0 & 0 
& 0 & 0 & 0 & 0 & 0 & 0 & 0 & 0 & 0\cr
 0 & 0 & 0 & 0 &-1 & 0 &-1 & 0 &-1 & 0 & 0 & 1 
& 0 & 0 & 0 & 0 & 0 & 0 & 0 & 0 & 0\cr
 0 & 0 & 0 & 0 & 0 & 0 & 0 & 0 & 1 & 0 & 0 & 0 
& 0 & 0 & 0 & 0 & 0 & 0 & 0 & 0 & 0\cr
 0 & 0 & 0 & 0 & 0 & 0 & 0 & 1 & 0 & 0 & 0 & 0 
& 0 & 0 & 0 & 0 & 0 & 0 & 0 & 0 & 0\cr
 0 & 0 & 0 & 0 & 0 & 0 & 1 & 0 & 0 & 0 & 0 & 0 
& 0 & 0 & 0 & 0 & 0 & 0 & 0 & 0 & 0\cr
 0 & 0 & 0 & 0 & 0 & 1 & 0 & 0 & 0 & 0 & 0 & 0 
& 0 & 0 & 0 & 0 & 0 & 0 & 0 & 0 & 0\cr
 0 & 0 & 0 & 0 & 1 & 0 & 0 & 0 & 0 & 0 & 0 & 0 
& 0 & 0 & 0 & 0 & 0 & 0 & 0 & 0 & 0\cr}
\right].\cr}}
The quiver gauge group consists of eight $U(1)$ factors, 
the diagonal $U(1)$ leaving the chiral multiplets $X^i_{M,M+w_i}$ 
invariant. Therefore, in solving the D-flatness constraints 
we have to divide by the effective group $G=U(1)^7$. 
We choose the seven independent $U(1)$ 
factors to correspond to the nodes $0,1,\ldots,6$ of the quiver. 
The charges of the twelve independent variables considered 
above are given by a $7\times 12$ matrix 
\eqn\Vmatrix{
V=\left[\matrix{
-1 &-1 & -1 & -1& -1 & 0 & 0 & 0 & 0 & 0 & 0 & 1\cr
 1 & 1 & 0 & 0 & 0 &-1 & 0 & 0 & 0 & 0 & 0 & 0\cr
 0 & 0 & 1 & 1 & 1 & 1 & -1 & 0 & 0 & 0 & 0 & 0\cr
 0 & 0 & 0 & 0 & 0 & 0 & 1 & -1 & 0 & 0 & 0 & 0\cr
 0 & 0 & 0 & 0 & 0 & 0 & 0 & 1 & -1 & 0 & 0 & 0\cr
 0 & 0 & 0 & 0 & 0 & 0 & 0 & 0 & 1 & -1 & 0 & 0\cr
 0 & 0 & 0 & 0 & 0 & 0 & 0 & 0 & 0 & 1 & -1 & 0\cr
}\right].}
The total charge matrix is obtained by concatenating $Q$ and $VU$ 
in a single $16\times 21$ matrix 
\eqn\tQmatrix{
{\widetilde Q}=\left[\matrix{Q\cr VU\cr}\right].}
Note that the quiver FI terms can also be included as an extra column 
$\left[\matrix{ {\widetilde Q} & \xi\cr}\right]$ where 
$\xi=\left[0^9,\xi_1,\ldots,\xi_7\right]^t$. 

The transpose of the kernel of ${\widetilde Q}$
gives the presentation of the
quiver moduli space as a toric variety. 
The columns of $\left(\hbox{Ker}{\widetilde Q}\right)^t$, 
interpreted as vectors 
in a linear space of appropriate dimension, generate 
the toric fan of the moduli space. In the present case, we obtain
\eqn\kermatrix{\eqalign{
&\left(\hbox{Ker}{\widetilde Q}\right)^t=\cr
&\left[\matrix{
-5& 0 & 0 & 2 & -1& -1& -1& -1& -1& -1& -1& -1& 1 & 1 
& 0 & 0 & 0 & 0 & 0 & 0 & 0 \cr
0 & 0 & 0 & -1& 0 & 0 & 0 & 0 & 0 & 0 & 0 & 0 & 0 & 0 
& 0 & 0 & 0 & 0 & 0 & 0 & 1\cr
-1& 1 & 0 & 0 & 0 & 0 & 0 & 0 & 0 & 0 & 0 & 0 & 0 & 0 & 0
& 0 & 0 & 0 & 0 & 0 & 0 \cr 
-1& 0 & 1 & 0 & 0 & 0 & 0 & 0 & 0 & 0 & 0 & 0 & 0 & 0 & 0
& 0 & 0 & 0 & 0 & 0 & 0 \cr 
4 & 0 & 0 & 0 & 1 & 0 & 1 & 0 & 0 & 0 & 0 & 0 & 0 & 0 & 0
& 0 & 0 & 0 & 0 & 0 & 0 \cr}
\right]\cr}}
Let ${\widetilde T}$ be the matrix obtained
after eliminating the redundant columns.
Given this matrix, we can find the associated charge
matrix by taking again the transpose of its kernel.
This yields 
\eqn\finalcharge{
\left[\matrix{
 1 & 1 & 1 & 0 & -4 & 1 & 0 & 0 \cr
 0 & 0 & 0 & 0 & -1 & -1 & 1 & 0\cr
 0 & 0 & 0 & 1 & 0 & -2 & 0 & 1 \cr}
\right].}
As stated before, the columns of this matrix correspond to 
homogeneous coordinates of the toric moduli space. We will denote 
them by $p_0,\ldots,p_7$. The presentation can be further simplified 
by using the middle charge vector to eliminate $p_6$ in terms 
of $p_4$ and $p_5$. Note that this is justified in a toric phase 
where $p_6$ is not allowed to vanish, since then it can be 
gauged away using the $\BC^*$ action. Assuming that we are
working in such a phase, we obtain a simplified charge matrix 
\eqn\simplified{
\left[\matrix{
 1 & 1 & 1 & 0 & -4 & 1 & 0 \cr
 0 & 0 & 0 & 1 & 0 & -2 & 1 \cr}
\right].}
We can set this data in more familiar form by permuting the columns
\eqn\simplifiedB{
\left[\matrix{
 1 & 1 & 1 & 1 &  0 & 0 & -4\cr
 0 & 0 & 0 & -2 &  1 & 1 & 0 \cr}
\right].}
This is easily recognizable as the toric data of the 
total space of the canonical line bundle $K_W$.
The first five columns, describe the smooth compact toric variety $W$
discussed in the beginning of the section.  
The last column corresponds to an extra
homogeneous coordinate representing the fiber
the canonical line bundle $K_W$. 

In order to justify such a interpretation, we have to make sure 
we are in the right toric phase, i.e., the disallowed locus is 
\toricB,\ while $p_7$ is allowed to vanish.  This turns out to be true
for a suitable interpretation of the Fayet--Iliopoulos terms arising from
the Gepner model construction.  Remember that these were $\xi_k = k \lambda$,
where $\lambda$ is a string-scale (known) mass parameter, and the origin
of the $k$ index is set arbitrarily (as we commented, the full spectrum cannot
in general be reproduced by a low energy supersymmetric theory).

The matrix ${\widetilde Q}$ with the associated FI terms 
can be set in canonical form \refs{\dgm,\partres} by invertible 
row operations. We record the result in appendix A. Note that 
the first two rows reproduce the charge matrix of $X$ \simplifiedB.\
The other rows can be used to eliminate redundant variables as 
explained below.
If all of $\xi_2$ through $\xi_8$ are positive, each row following the
first two can be used to eliminate a unique variable (the one appearing
with a negative charge), resulting in \simplified.  (The last row may
appear problematic as $\xi_7$ appears with the wrong sign, but if one
writes out the immediately preceding
moment map equations $|p_i|^2-|p_{i+1}|^2=-\xi$,
one sees that so many negative FI terms (including $-\xi_7$) have 
appeared
that the final variable is guaranteed to be nonzero).

Thus we have obtained the expected result---a smooth 
noncompact toric fivefold $X$ with vanishing canonical 
class. 

We now determine in a similar fashion the tautological line bundles. 
The construction explained in the previous section can be implemented
in practice by certain simple modifications of the 
quiver diagram. 
More precisely, we will 
consider a different gauge theory, obtained from the previous one 
by adding an extra chiral multiplet, corresponding to an extra 
leg in the diagram. The extra leg is attached to a single node,
resulting in a multiplet charged under the 
corresponding $U(1)$ factor. Therefore 
we obtain eight distinct quiver theories which can be labeled by 
the charge vector $v_M$ of the extra multiplet. 
Recall that we have fixed the gauge group 
to be
$G=U(1)_0\times U(1)_1\times \ldots \times U(1)_6$. 
We have 
\eqn\extracharge{
v_M^i=\delta_{M}^i,\qquad M =0,\ldots,7,
\qquad i=0,\ldots,6.}

Each quiver theory has a moduli space, whose toric presentation can 
be identified (in a certain phase) to the total space of a certain 
line bundle 
over the variety $X$ determined above. We claim that the line 
bundles obtained this way are precisely the K-theory generators 
$R_k$ introduced in section two. The precise correspondence between 
$R_k$ and the charge vectors $v_M$ is given by $k=M+1$.

In order to prove this, we 
proceed as before. The extra multiplet $\psi$ corresponds to an
extra variable which is not related to $X^1_{01},\ldots,X^1_{70}$. 
Therefore, the cone of monomials $M_+^\prime$ can be obtained 
by embedding $M_+$ in a hyperplane $\BR^{12}\subset \BR^{13}$ and 
adding an 
extra generator corresponding to the normal direction. 
Then, the dual cone $N_+^\prime$ is characterized by the 
augmented matrix 
\eqn\augTmatrix{
T^\prime=\left[\matrix{
T & 0\cr  0 & 1\cr}\right].}
Similarly, $Q^\prime =\left(\hbox{Ker}T^\prime\right)^t$ and $U^\prime$ 
can be easily related to $Q,U$
\eqn\augQUmatrix{
Q^\prime = \left[\matrix{Q &0\cr}\right],\qquad 
U^\prime = \left[\matrix{U &0\cr 0 &1\cr}\right].}
The matrix $V$ is augmented by the charge vector of the 
extra multiplet\foot{Note that we have in fact eight different 
charge matrices $V^\prime$ which should be labeled by an index $M$. 
In order to keep the notation simple, we will not write down this 
extra index explicitly.}
\eqn\augVmatrix{
V^\prime = \left[\matrix{V&v_M\cr}\right]}
resulting in a total charge matrix 
\eqn\augtQ{
{\widetilde Q}^{\prime}=\left[\matrix{Q&0\cr 0&v_M\cr}\right].}
This is an important change, since the matrix ${\widetilde Q}^{\prime}$
determines the new moduli space. 
 
Following the general algorithm, the next step is to determine the 
transpose of the kernel of ${\widetilde Q}^{\prime}$.
This can be done by a straightforward computation for a generic 
vector $v_M$. We do not record the result here for reasons of space. 
After eliminating the redundant column, we are left with a matrix 
${\widetilde T}^{\prime}$ as before. 
At the last stage, we can determine the charge matrix of the resulting 
toric variety by taking the transpose of the kernel of 
${\widetilde T}^{\prime}$. Let us carry out this procedure explicitly 
for 
\eqn\explict{
v_0=\left[\matrix{1&0&0&0&0&0&0}\right]^t.}
The matrix ${\widetilde T}^{\prime}$ reads in this case 
\eqn\redT{
{\widetilde T}^{\prime}=\left[\matrix{
-5 & 0 & 0 & 2 & -1 & -1 & 1 & 0 & 1 & 0 & 0\cr
 -1 & 1 & 0 & 0 & 0 & 0 & 0 & 0 & 0 & 0 & 0\cr
 -1 & 0 & 1 & 0 & 0 & 0 & 0 & 0 & 0 & 0 & 0\cr
 4  & 0 & 0 & 0 & 1 & 1 & 0 & 1 & 0 & 0 & 0\cr
 4  & 0 & 0 &-1 & 0 & 1 & -1& 0 & 0 & 0 & 1\cr
 0 & 0 & 0 & -1 & 0 & 0 & 0 & 0 & 0 & 1 & 0\cr}
\right].}
The associated charge matrix is 
\eqn\augcharge{
\left[\matrix{
 1 & 1 & 1 & 0 & -4& 0 & 1 & 0 & 0 & 0 & -3\cr
 0 & 0 & 0 & 1 & 0 & 0 & -2& 0 & 0 & 1 & -1\cr
 0 & 0 & 0 & 0 & 1 & -1& 0 & 0 & 0 & 0 & 1\cr
 0 & 0 & 0 & 0 & 1 & 0 & 1 & -1& 0 & 0 & 1\cr
 0 & 0 & 0 & 0 & 0 & 0 & 1 & 0 & -1& 0 & 1\cr}\right].}
Note again that the last three charge vectors can be used to 
gauge away $p_5,p_7,p_8$ in a phase where they are not allowed
to vanish. Assuming that we are working in such a phase,
and permuting again the columns, 
the charge matrix can be rewritten as 
\eqn\augsimplf{
\left[\matrix{
 1 & 1 & 1 & 1 & 0 & 0 & -4& -3\cr
 0 & 0 & 0 & -2& 1 & 1 & 0 & -1\cr}\right].}
The first seven columns constitute the toric data of $X$ 
determined before. By adding the last column, we find that the 
new toric variety $X_1$ can be interpreted in a certain phase 
as the total space of a line bundle $R_1$ over $X$. The fiber 
is described by the last homogeneous variable $p_{10}$, so this 
interpretation is justified if $p_{10}$ is allowed to vanish. 
The existence of a suitable phase can be proved by a direct 
computation of the FI terms, as before. 
In this case, the line bundle $R_1$ is determined by the entries 
in the last column to be 
\eqn\lbA{
R_1={\cal O}(-3H-L),}
where $(H,L)$ generate the Picard group of the resolved 
$\WP^{1,1,2,2,2}$. 
The bundle in the r.h.s. of \lbA\ is pulled back to $X$. 

Proceeding similarly we can determine all eight line bundles 
$R_k$, $k=1,\ldots,8$
\eqn\seqlb{
\matrix{
&R_1=\CO(-3H-L)\hfill & &  R_2=\CO(-3H)\hfill\cr
& & & \cr
&R_3=\CO(-2H-L)\hfill & & R_4=\CO(-2H)\hfill\cr
& & & \cr
&R_5=\CO(-H-L)\hfill & & R_6=\CO(-H)\hfill\cr
& & & \cr
&R_7=\CO(-L)\hfill & & R_8=\CO.\hfill\cr
}}

\subsec{A dual basis}

We now test the main conjecture in this paper by determining 
a set of dual K-theory classes $S_l$. Strictly speaking, we will 
only determine the Chern characters $\ch(S_l)$  
so that the following orthogonality relation holds
\eqn\Korth{
\int_X\ch(R_k)\ch(S_l)\hbox{Td}(X) =\delta_{kl}.}
The result is conveniently expressed in terms of a K-theory 
class $S$ defined by 
\eqn\Sclass{\eqalign{
& \ch(S)=3-(2H-L)+{1\over 12}(2H-L)^3\cr}}
and its conjugate ${\overline S}$. 
Then the eight dual classes are given by 
\eqn\dualK{
\matrix{
&S_1=\CO(-H+L)\hfill & & S_2=-\CO(-H+2L)\hfill\cr
& & & \cr
&S_3=-S\hfill & & S_4=S\otimes\CO(L)\hfill\cr
& & & \cr
&S_5={\overline S}\otimes\CO(-H)\hfill & & 
S_6=-{\overline S}\otimes\CO(-H+L)\hfill\cr
& & & \cr
&S_7=-\CO(-L)\hfill & & S_8=\CO.\hfill\cr}}
It can be checked by a direct computation that the classes 
$R_k, S_l$ defined above satisfy \Rintform--\Sintform\ and \Korth.

In order to establish a relation to Gepner model boundary states, 
we restrict the classes $S_l$ to the Calabi--Yau hypersurface 
$M$. Let $V_l$ denote the restriction of $S_l$ to $M$. 
Using the intersection relations \intD,\ it is straightforward 
to compute the Chern characters $\ch(V_l)$ 
\eqn\reschern{
\eqalign{
&\ch(V_1)=1-H+L+2l+{2\over 3}\cr
&\ch(V_2)=-1+H-2L+4h-2l-{8\over 3}\cr
&\ch(V_3)=-3+2H-L-{4\over 3}\cr
&\ch(V_4)=3-2H+4L-8h+{4\over 3}\cr
&\ch(V_5)=3-H-L-2l+{2\over 3}\cr
&\ch(V_6)=-3+H-2L+4h+2l+{4\over 3}\cr
&\ch(V_7)=-1+L\cr
&\ch(V_8)=1.\cr}}

In the remaining part of this section, we will compare \reschern\
to the topological invariants of the fractional 
branes, finding a precise agreement. 

\subsec{Fractional branes}

The geometric interpretation of Gepner model boundary states 
for Calabi--Yau models has been considered in 
\refs{\bdlr,\dr,\lerche,\schei}. In particular, the present 
two-parameter model has been studied in \lerche. 
Here we determine a complete list of K-theory classes 
in $K^0(M)$ corresponding to the fractional branes. 

Let us give some background on the special K\"ahler geometry of 
$\WP^{1,1,2,2,2}$ \twopar. We adopt the conventions of \lerche\
for the basis of periods. The K\"ahler moduli space is parameterized
near the large radius limit by 
\eqn\kahlerA{
J = t_1H + t_2 L.}
The asymptotic expression of the prepotential (ignoring exponentially 
small corrections) is 
\eqn\perA{
F=-{4\over 3}t_1^3-2t_1^2t_2+{7\over 3}t_1+t_2}
which yields the following vector of periods
\eqn\perB{
\Pi(t)=\left[\matrix{
{4\over 3}t_1^3+2t_1^2t_2+{7\over 3}t_1+t_2\cr
-4t_1^2-4t_1t_2+{7\over 3}\cr
-2t_1^2+1\cr
1\cr
t_1\cr
t_2\cr}\right].}
The monodromy matrix corresponding to the $\BZ_8$ quantum symmetry of 
the Gepner model is \lerche
\eqn\perC{
A=\left[\matrix{-1 & 1 & 0 & 0 & 0 & 0\cr
                {3\over 2}&{3\over 2}&0&0&-{1\over 2}&-{1\over 2}\cr
                 1 & 0 & 1 & 0 & 0 & 0\cr
                 1 & 0 & 0 & 0 & 0 & 0\cr
                -{1\over 4}& 0 & {1\over 2}& 0 & {1\over 4}& 0\cr
                {1\over 4}&{3\over 4}&-{1\over 2}&{1\over 2}&
                -{1\over 4}&{1\over 4}\cr}\right]}
We label BPS states by a six dimensional charge vector 
$n=(n_6,n_4^1,n_4^2,n_0,n_2^1,n_2^2)$, with central charge 
\eqn\chchargeA{\eqalign{
Z(n)&=n\cdot\Pi(t)\cr
&={4\over 3}n_6t_1^3+2n_6t_1^2t_2-(4n_4^1+2n_4^2)t_1^2\cr
&\ -4n_4^1t_1t_2+(n_2^1+{7\over 3}n_6)t_1+(n_2^2+n_6)t_2\cr
&\ +n_0+{7\over 3}n_4^1+n_4^2.\cr}}
This is to be compared with the central charge of a D-brane
configuration described by a bundle $V\rightarrow M$ 
\eqn\chchargeB{\eqalign{
Z(V)&=\int_Xe^{-(t_1H+t_2L)}\ch(V)\sqrt{\td(M)}\cr
&=\int_Xe^{-(t_1H+t_2L)}\ch(V)\left(1+{c_2(M)\over 24}\right).\cr}}
By identifying \chchargeA\ and \chchargeB\ we obtain the 
conversion formulae
\eqn\conv{\eqalign{
&\ch_0(V) = -n_6\cr
&\ch_1(V)=-n_4^1H-n_4^2L\cr
&\ch_2(V)=-n_2^1h-n_2^2l\cr
&\ch_3(V)=n_0+{14\over 3}n_4^1+2n_4^2.\cr}}

The fractional branes generically correspond to Gepner model 
boundary states with $L=(0,0,0,0,0)$ and they form an orbit 
of the $\BZ_8$ quantum symmetry. Typically, one of these states 
corresponds to the pure D6-brane wrapping the CY hypersurface
$M$, with charge 
vector\foot{The label $n_8$ is for further convenience.}
\eqn\puresix{
n_8=(-1,0,0,0,0,0).}
The other charge vectors are obtained by multiplying by $A^{-1}$ to the 
right. We obtain the following charges
\eqn\charges{\eqalign{
& n_1=(-1,1,-1,-2,0,-2)\cr
& n_2=(1,-1,2,-2,-4,2)\cr
& n_3=(3,-2,1,6,0,0)\cr
& n_4=(-3,2,-4,0,8,0)\cr
& n_5=(-3,1,1,-6,0,2)\cr
& n_6=(3,-1,2,2,-4,-2)\cr
& n_7=(1,0,-1,2,0,0).\cr}}
Using \conv,\ we can now determine the topological invariants of the 
K-theory classes associated to these BPS states. 
A straightforward computation shows that they are in precise agreement 
with \reschern.\ This proves the claim. 

\newsec{Conclusions}

In this paper we have continued the investigation of D-branes
in Gepner models and their geometric counterparts 
initiated in \refs{\rs,\bdlr}.  One of the central problems
in this area is to find a general geometric interpretation for Gepner 
model B boundary states in terms of holomorphic objects. 
So far, this question has been answered in several particular
models using mirror symmetry techniques, but no general 
picture had been found. 

The present work fills this gap by proposing a simple construction 
of the K theory elements associated to boundary states. The power 
of this new construction is that it does not 
make use of mirror symmetry results. The approach is inspired by 
the description of fractional branes in lower dimensional 
orbifold models \refs{\dm,\dgm}. In these cases, the geometric 
interpretation of fractional branes essentially reduces to 
the celebrated McKay correspondence
\refs{\mckay,\reid,\reidbourbaki}. The latter establishes
a duality between orbifold (equivariant) K theory and the K theory of
the resolved space. Given the K theoretic interpretation of branes
\refs{\MM,\DK}, this is precisely our problem, formulated in a
slightly different language.

We argued that our problem, although superficially different, is
essentially a higher dimensional version of the McKay correspondence,
less well studied in the mathematical literature.  This follows
naturally from the Landau-Ginzburg orbifold description of the Gepner
model, which is a familiar point of view in the linear sigma model
approach of \witten.  By shifting perspective from the conformal field
theory approach of \refs{\rs} to the linear sigma model, we obtain a
simple and efficient description of boundary states as fractional
branes in a ${\BC}^5/\Gamma$ orbifold.  Indeed, if we keep in mind
that only quantities which are topological or protected in some way
should be directly computable in the UV, we can derive the Gepner
model results we use, namely the $\CN=1$ effective Lagrangian which
describes combinations of rational boundary states, from the LG
orbifold description.

It is well known from the work of \refs{\dg,\dgm} that orbifold
boundary states are described by quiver gauge theories whose moduli
space typically reproduces the orbifold resolution. Moreover, in this
picture, one can naturally define a preferred set of K theory
generators of the resolution -- the tautological line bundles $R_k$
\refs{\kronone,\kronnak,\itonakajima}. 
These are dual to another set of K theory generators $S_k$ supported 
on the compact exceptional locus of the blownup singularity. 
For ${\BC}^5/\Gamma$ orbifolds, the resolved 
space is a noncompact Calabi-Yau variety isomorphic to the
total space of the canonical line bundle of a weighted projective 
space. 

The main result of the present work establishes a direct correspondence 
between fractional branes and the dual set of K theory generators 
$S_k$, which can be thought of as classes in the K theory of 
the weighted projective space. This shows that the Landau-Ginzburg 
orbifold is intimately related to the geometry of 
the ambient toric variety which contains the Calabi-Yau 
hypersurface. 

In practice we have developed a systematic toric algorithm for 
determining the tautological line bundles $R_k$, starting from 
the quiver diagram. The basis of fractional branes is then
determined by inverting the K theoretic intersection 
pairing on the ambient toric variety.  Finally, these are restricted
to the hypersurface to obtain the classes of Gepner model boundary states.

One surprising aspect of this is that the intersection form on the
ambient variety is different from the intersection form on the
Calabi-Yau hypersurface itself, which reproduces the CFT results.
Although this seems to be a natural aspect of the whole picture, the
objects $S_k$ defined above are not really physical, and neither is
their intersection form: only their restrictions to the CY manifold
are physical.  Nevertheless, one gets correct results by working with
the unphysical $S_k$'s.  This seems to be another manifestation of the
decoupling between D-flatness conditions (which define the ambient
toric space) and F-flatness conditions (which define the CY), but
remains somewhat mysterious.

We feel that the ideas presented here are the beginnings of a
satisfactory picture both of the world-sheet and space-time
interpretations of the rational boundary states, but there is much
work still to do in this direction.  It would be very useful to
further develop the linear sigma model technology and complete the
computations of all the ``topological'' results: spectrum,
superpotentials, and D-flatness conditions.  The conditions under
which a low energy theory treatment of bound states of branes is
valid, and what must replace it in general, remain to be explored.

The central point for further development is that, as is clear from
\refs{\itonakajima,\bkr}\ and the other references, the generalized
McKay correspondence can be used not just to compute K theory classes
but to determine explicit holomorphic objects (sheaves or complexes
of sheaves) representing the fractional branes.  
Let us recall what appears to be the central lesson of \noncompact:
the detailed $\CN=1$ world-volume theories describing combinations and bound
states of fractional branes on a CY, have a direct correspondence to
natural mathematical constructions of stable holomorphic objects.
For $\BC^3/\BZ_3$, the core of this was the identification of
Beilinson's construction of sheaves on $\P^2$ within the $\BZ_3$
quiver gauge theory.  It is precisely this identification which is
generalized in the works \refs{\itonakajima,\bkr}.

Thus, we can hope to generalize this lesson to fairly general CY's.
The simple conjecture (an oversimplification, in ways described in
detail in \refs{\noncompact,\category} and elsewhere) which can guide
further developments is that moduli spaces of $\CN=1$ theories
describing bound states of fractional branes naturally correspond to
moduli spaces of stable coherent sheaves on the CY, and furthermore
the internal structure (F and D flatness conditions) of these theories
corresponds to natural mathematical descriptions of the moduli space.
If so, the constructions we described could then eventually lead to a
construction of all sheaves on a CY which can be obtained by
restriction from the ambient toric variety and then deformation.  This
is a sublattice of finite index in the K-theory of the CY (for example
the D$0$ did not appear on the quintic) but still gives a very large
subset of the possibilities.

The most interesting physical application of these results may
be to constructing type \I\ compactifications on Calabi--Yau
manifolds, by finding combinations of branes and orientifolds which
cancel tadpoles and anomalies, and finding the supersymmetric vacua of
these theories.  A geometric study of type \I-heterotic duality should
then be possible; indeed the $\BC^3/\Gamma$ results should already be
quite useful for this purpose.  A natural next step for both flat
space and Landau--Ginzburg orbifolds would be to show that the tadpole
cancellation conditions are equivalent to the familiar anomaly
cancellation conditions for heterotic strings on the corresponding
bundles.

\medskip

We would like to thank S. Katz, E. Martinec, G. Moore, A. Sen, E. Witten,
and especially D. Morrison for helpful discussions and comments.

This research was supported in part by DOE grant DE-FG02-96ER40959
and by DOE grant DE-FG02-90ER40542.

\listrefs
\vfill\eject

\appendix{A}{Fayet-Iliopoulos Terms}

\nopagenumbers
\newbox\rotright
\hskip100pt
\setbox\rotright=
\hbox{$ 
\left [\matrix{
1 & 1 & 1 & 1 & 1 & 0 & -4 & 0 & 0 & 0 & 0 & 0 & 0 & 0 & 1 & 0 & 0 & 0 
& 0 & 0 & 0 & 0 & 0 & 
\xi_3+\xi_4+2\xi_5+2\xi_6+\cr
 & & & & & & & & & & & & & & & & & & & & &  &  &  {3\xi_7+3\xi_8}\cr
0 & 0 & 0 & 0 & 0 & 1 & 0 & 0 & 0 & 0 & 0 & 0 & 0 & 0 & -2 
& 0 & 0 & 0 & 0 & 0 & 0 & 0 & 1 & \xi_2+\xi_4+\xi_6+\xi_8\cr
0 & 0 & 0 & 0 & 0 & 0 & 1 & -1 & 0 & 0 & 0 & 0 & 0 & 0 & 0 & 0 & 0 & 0 
& 0 & 0 & 0 & 0 & 0 & -\xi_8\cr
0 & 0 & 0 & 0 & 0 & 0 & 1 & 0 & 0 & 0 & 0 & 0 & 0 & 0 & 1 & 0 & -1 
& 0 & 0 & 0 & 0 & 0 & 0 & -\xi_6 -\xi_8\cr
0 & 0 & 0 & 0 & 0 & 0 & 0 & 1 & -1 & 0 & 0 & 0 & 0 & 0 & 0 & 0 & 0 & 0 
& 0 & 0 & 0 & 0 & 0 & -\xi_7\cr
0 & 0 & 0 & 0 & 0 & 0 & 0 & 0 & 1 & -1 & 0 & 0 & 0 & 0 & 0 & 0 
& 0 & 0 & 0 & 0 & 0 & 0 & 0 & -\xi_6\cr
0 & 0 & 0 & 0 & 0 & 0 & 0 & 0 & 0 & 1 & -1 & 0 & 0 & 0 & 0 & 0 & 0 
& 0 & 0 & 0 & 0 & 0 & 0 & -\xi_5\cr
0 & 0 & 0 & 0 & 0 & 0 & 0 & 0 & 0 & 0 & 1 & -1 & 0 & 0 & 0 & 0 & 0 
& 0 & 0 & 0 & 0 & 0 & 0 & -\xi_4\cr
0 & 0 & 0 & 0 & 0 & 0 & 0 & 0 & 0 & 0 & 0 & 1 & -1 
& 0 & 0 & 0 & 0 & 0 & 0 & 0 & 0 & 0 & 0 & -\xi_3\cr
0 & 0 & 0 & 0 & 0 & 0 & 0 & 0 & 0 & 0 & 0 & 0 & 1 & -1 
& 0 & 0 & 0 & 0 & 0 & 0 & 0 & 0 & 0 & -\xi_2\cr
0 & 0 & 0 & 0 & 0 & 0 & 0 & 0 & 0 & 0 & 0 & 0 & 0 & 0 & 0 & 0 & 1 & -1 
& 0 & 0 & 0 & 0 & 0 & -\xi_4\cr
0 & 0 & 0 & 0 & 0 & 0 & 0 & 0 & 0 & 0 & 0 & 0 & 0 & 0 & 0 & 0 & 0 & 1 
& -1 & 0 & 0 & 0 & 0 & -\xi_7\cr
0 & 0 & 0 & 0 & 0 & 0 & 0 & 0 & 0 & 0 & 0 & 0 & 0 & 0 & 0 & 0 & 0 & 0 
& 1 & -1 & 0 & 0 & 0 & -\xi_2\cr
0 & 0 & 0 & 0 & 0 & 0 & 0 & 0 & 0 & 0 & 0 & 0 & 0 & 0 & 0 & 0 & 0 & 0 
& 0 & 1 & -1 & 0 & 0 & -\xi_5\cr
0 & 0 & 0 & 0 & 0 & 0 & 0 & 0 & 0 & 0 & 0 & 0 & 0 & 0 & 0 & 0& 0 & 0 
& 0 & 0 & 1 & -1 & 0 & \xi_7\cr}
\right]
$}
\rotr{\rotright}

\end